\documentclass[aps,pre,twocolumn,superscriptaddress,amssymb]{revtex4-2}
\usepackage{physics}
\usepackage{bm}
\usepackage{appendix}
\usepackage{mathtools}
\usepackage{comment}

\newcommand{\argmax}{\mathop{\rm arg~max}\limits}
\newcommand{\argmin}{\mathop{\rm arg~min}\limits}

\begin{document}

\title{Effect of global shrinkage parameter of horseshoe prior in compressed sensing }
\author{Yasushi Nagano}
\address{
 Graduate School of Arts and Sciences,
 The University of Tokyo,
 Komaba, Meguro-ku, Tokyo 153-8902, Japan
}

\author{Koji Hukushima}
\address{
 Graduate School of Arts and Sciences,
 The University of Tokyo,
 Komaba, Meguro-ku, Tokyo 153-8902, Japan
}
\address{
Komaba Institute for Science, The
University of Tokyo, 3-8-1 Komaba, Meguro-ku, Tokyo 153-8902, Japan
}
\begin{abstract}
 In sparse signal processing, this study investigates the effect of the global shrinkage parameter $\tau$ of a horseshoe prior, one of the global-local shrinkage prior, on the linear regression. Statistical mechanics methods are employed to examine the accuracy of signal estimation. A phase diagram of successful and failure of signal recovery in noise-less compressed sensing with varying $\tau$ is discussed from the viewpoint of dynamic characterization of the approximate message passing as a solving algorithm and static characterization of the free-energy landscape. It is found that there exists a parameter region where the approximate message passing algorithm can hardly recover the true signal, even though the true signal is locally stable. The analysis of the free-energy landscape also provides important insight into the optimal choice of $\tau$. 
\end{abstract}
\date{\today}

\maketitle

\section{Introduction}
Recently, sparse signal estimation has played an important role in signal processing, machine learning, image processing, and communication. Sparse signals have a very small number of non-zero components, and the purpose of sparse signal estimation is to accurately estimate the non-zero components. It has attracted much attention because of its high accuracy and efficiency in applications such as signal reconstruction. In the field of physics, the technique has been applied to experimental measurements, and theoretical performance evaluation of the efficiency of solving algorithms has been studied using statistical mechanics methods. 

There are various methods for estimating a sparse signal, among which the use of a global-local shrinkage prior has been extensively studied. This prior is characterized by the unique structure of the variance parameters within a normal distribution, composed of a product of two distinct types: a global variance parameter and a local variance parameter. The global variance parameter is independent of the index, representing a universal attribute, whereas the local variance parameter is dependent on the index, capturing individual characteristics. This family of prior distributions includes well-known priors used in sparse signal estimation, such as the Laplace prior and the Automatic Relevance Determination prior. Among them, the horseshoe prior is one of the most promising in this family\cite{tibshirani1996regression,neal2012bayesian,carvalho2009handling,carvalho2010horseshoe}. Theoretical analysis of the properties of sparse signal estimation using the horseshoe prior has been a topic of interesting research. It is shown that the horseshoe prior asymptotically achieves the minimax rate,  known as the lower bound of sparse signal estimation with respect to the estimated $\ell_2$ risk\cite{donoho1992maximum}, and this result is being extended to the general global-local shrinkage parameter\cite{van2014horseshoe,van2017adaptive,van2016conditions,ghosh2016asymptotic}. The horseshoe prior is also known to have asymptotic Bayes optimality under sparsity with respect to the risk of variable selection, which is an advantage over the widely-used LASSO, the case for the Laplace prior\cite{bogdan2011asymptotic,datta2013asymptotic}. 

In constructing the global-local shrinkage prior, it is crucial to select an appropriate prior for the local shrinkage parameters and to adjust the global shrinkage parameter correspondingly. Several prior distributions have been proposed for the local shrinkage parameter, with particular emphasis on the utility of heavy-tailed distributions. There are also  practical methods for determining the appropriate global shrinkage parameters, such as the maximum marginal likelihood estimator and the fully Bayesian approach\cite{bhadra2019lasso}. However, the theoretical understanding of the sparse linear regression with the horseshoe prior has not yet been fully explored\cite{bhadra2019prediction}. To be specific, the effect of the choice of the global shrinkage parameter on the estimation accuracy in sparse linear regression remains unclear. 

One of the key challenges in sparse modeling is compressed sensing, a subfield of sparse linear regression. Compressed sensing is a technique for precisely inferring an unknown vector for which sparsity is assumed from a smaller number of observations than the dimension of the vector. Therefore, it is extremely important to develop algorithms that can effectively estimate the signal with a limited number of observations. 
The Belief Propagation (BP) algorithm, or its approximation, Approximate Message Passing (AMP), is an important algorithm for solving the problem of compressed sensing\cite{donoho2010message}.

The typical performance of the algorithms for compressed sensing  has been widely discussed in the field of statistical physics using the replica method. Analysis of the $l_1$-norm regularization has found that linear stability of the replica symmetric (RS) free energy in the vicinity of the true signal is an important criterion for successful signal restoration\cite{kabashima2009typical,donoho2009observed}. 
 It is also valuable to be able to track the dynamics of the algorithm through State Evolution (SE), which describes the macroscopic temporal evolution of AMP\cite{donoho2010message_2}. Under appropriate assumptions about the threshold function, a formal equivalence can be shown between the time evolution of SE and the saddle-point equation for the RS free energy. In particular,  the fixed points of SE for AMP coincide with the saddle point of the RS free energy.
 In the Bayesian optimal setting with the Gaussian-Bernoulli distribution,  the free-energy landscape as a function of the mean squared error reveals the potential existence of spurious local solutions in addition to the true solution. This indicates that solving algorithms such as BP and AMP may fail to recover the true signal even when the true signal satisfies the linear stability condition, implying the existence of an algorithmic transition point at which the local fixed point of SE vanishes\cite{krzakala2012statistical,krzakala2012probabilistic}. 

These statistical physics results are obtained by taking the expected value of the free energy with respect to the randomness of the problem. The method of computation used is mainly based on the replica method of spin glass theory, which requires close attention to its mathematical justification. The exactness of the method is known for the mean squared error in the case of Gaussian i.i.d. randomness, which is the generative model addressed in this study\cite{reeves2016replica}.
 In this paper, extending our previous work\cite{nagano2023phase} focusing on the local shrinkage parameter, we study the role of the global shrinkage parameter in compressed sensing with the horseshoe prior using a statistical mechanics method based on the replica method and demonstrate that the signal recovery performance is characterized by a complex free-energy structure. By appropriately adjusting the global shrinkage parameter, it is found that the signal recovery limit with the horseshoe prior is competitive with that of the Bayes optimal method, even though it does not assume knowledge of the generative distribution of the true signal. As a byproduct, we have also been able to give an interpretation of damping in the solving algorithms in terms of the free-energy landscape. This result highlights the importance of considering the free energy landscape as a multivariable function.

The remainder of this paper is organized as follows. In Sec.~\ref{sec:2}, we describe the three setups based on the theoretical performance evaluation of the compressed sensing with the horseshoe prior, the approximate message passing as a solving algorithm, the state evolution characterizing the macroscopic behavior of the algorithm, and the formulation of  the statistical mechanics with the replica method.  In Sec.~\ref{sec:3}, we theoretically analyze the dynamical behavior of the solving algorithm and derive the corresponding phase diagram. In Sec.~\ref{sec:4}, the free-energy landscape is introduced to give a static interpretation to the phase diagram obtained in Sec.~\ref{sec:3}, and the effects of damping in the algorithm are discussed in terms of the free-energy landscape. Finally, Sec.~\ref{sec:5} contains a summary and discussion of our results. Some details of the calculations are presented in Appendices A, B, C, and D.

\section{Problem settings and Formulation}
\label{sec:2}
\subsection{Compressed sensing with horseshoe prior and Approximate Message Passing}
Compressed sensing is usually defined as an optimization problem of a normalization term $\mathcal{H}(\bm w)$ with a constraint: 
\begin{equation}
\hat {\bm w} \coloneqq \argmin_{\bm w}\mathcal{H}(\bm w) ~ \mathrm{s.t.}~\bm y = \bm X\bm w, 
\end{equation}
where $\hat{\bm w}\in\mathbb{R}^N$ is an estimator for the true signal $\bm w_0\in\mathbb{R}^N$, $\bm X\in\mathbb{R}^{M\cross N}$ is a design matrix, and $\bm y\in\mathbb{R}^M$ is observations. In the noiseless case, $\bm y$ is obtained by the product of the design matrix $\bm X$ and the true signal $\bm w_0$ as 
\[
\bm y = \bm X\bm w_0, 
\]
where the fraction of non-zero element in $\bm{w}_0$ is denoted by $\rho$. 
The mean-squared error (MSE), which is often used to determine the accuracy of the estimated signal, is expressed using the $\ell_2$ norm as 
\begin{equation}
   \mathrm{MSE} = \frac{1}{N}\left|\hat{\bm w}-{\bm w}_0\right|_2^2. 
\end{equation}

In Bayesian statics, the optimization problem is interpreted as a Maximum a Posterior (MAP) estimation with the likelihood $\delta(\bm y-\bm X\bm w)$ and a prior distribution $p(\bm w)$. Clearly, the optimization problem and the MAP estimation are connected by 
\[
\mathcal{H}(\bm w) = -\ln p(\bm w).
\]
In this study, the horseshoe prior is taken as the prior distribution, where there are two different parameters, a local and a global shrinkage parameter. In our model, the local shrinkage parameters $\{\lambda_i|i=1,..., N\}$ are determined by a MAP estimation, and the global shrinkage parameter $\tau$ is given as a hyperparameter. The detail of the horseshoe prior is introduced later.

\begin{figure}
    \centering
    \includegraphics[width=1.0\linewidth]{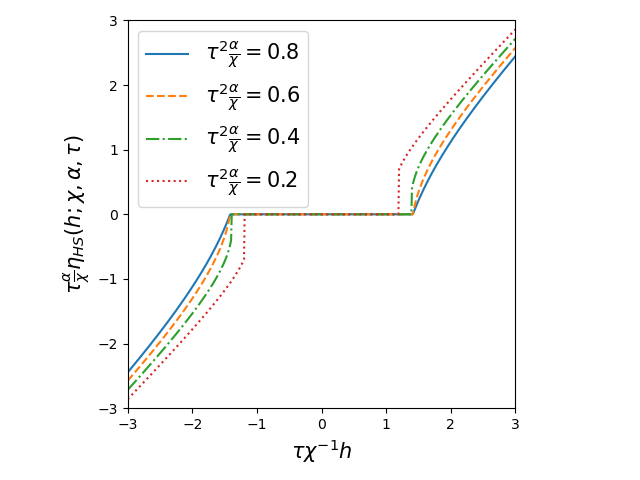}
    \caption{
    Threshold function $\eta_{\mathrm{HS}}$ as a function of the scaling parameter $\tau\chi^{-1}h$ for some value of the parameter $\tau^2\alpha/\chi$. 
    }
    \label{fig:threshold}
\end{figure}

For $X_{ij}\in\{\frac{1}{\sqrt{N}},-\frac{1}{\sqrt{N}}\}$, Approximate Message Passing (AMP) is known to simplify BP in the large system size limit $N\rightarrow\infty$ with constant $\alpha \coloneqq \frac{M}{N}$, generally given by a set of iterative equations with 
\begin{align}
    \label{eqn:AMP}
    \begin{split}
 h^t_i &= \alpha w^t_i+\sum_b X_{bi}z^t_b,\\
    w^{t+1}_i &= \eta(h^t_i;\chi_t), \\
    z^{t+1}_a &= y_a-\sum_{j}X_{aj}w^{t+1}_{j}+z^t_a\frac{1}{N}\sum_i\eta'(h^t_i;\chi_t),\\
    \chi_{t+1} &= \chi_t\frac{1}{N}\sum_i\eta'(h^t_i;\chi_t). 
    \end{split}
\end{align}
The threshold function $\eta(h;\tau)$ in Eqs.~(\ref{eqn:AMP}) is defined for the horseshoe prior as
\begin{equation}
    \eta_{\mathrm{HS}}(h;\chi,\alpha,\tau) = \frac{\chi^{-1}h}{\chi^{-1}\alpha+\tau^{-2}(\lambda^*)^{-2}},\nonumber
\end{equation}
where 
\begin{equation}
    \lambda^* = \argmax_{\lambda}\frac{\chi^{-2}h^2}{2(\chi^{-1}\alpha+\tau^{-2}\lambda^{-2})}-\ln(1+\lambda^2).
    \label{eqn:t_function}
\end{equation}
When there is no risk of confusion, we abbreviate $\eta(h;\chi,\alpha,\tau)$ as $\eta(h;\chi)$. Fig.~\ref{fig:threshold} shows $\eta_{\mathrm{HS}}$ for some parameters $\tau^2\alpha/\chi$. 
When $|h|$ is small enough, $\eta_{\mathrm{HS}}$ returns $0$, which means the elimination of small coefficients. For $\tau^2\frac{\alpha}{\chi}\geq0.5$, $\eta_{\mathrm{HS}}$ is a continuous function, and  $\eta_{\mathrm{HS}}=0$ when $\tau \chi^{-1}|h| < \sqrt{2}$. For large $|h|$, the functions asymptotically reach a straight line with $y=x$, meaning no shrinkage for large inputs. Meanwhile, when $\tau^2\frac{\alpha}{\chi}<0.5$,  $\eta$ is a discontinuous function, thus AMP is not valid there.

\subsection{State Evolution}
The AMP algorithm, shown in Eq.~(\ref{eqn:AMP}), refines the estimation of signals by iterative calculations, starting from an appropriate initial condition. It is known that the estimation performance of AMP is described by the deterministic evolution of a few parameters, called state evolution (SE)\cite{donoho2010message_2}. To be more specific, the state evolution is an iterative equation with ``time" $t$ for the MSE, denoted $\sigma^2_t$, and the variance parameter $\chi_t$ of the estimation of AMP, given by 
\begin{equation}
\begin{split}
    \sigma^2_{t+1} &= \mathbb{E}_{w_0,\xi}\qty[\qty(\eta(\alpha w_0+\sqrt{\alpha\sigma^2_t}\xi;\chi_t)-w_0)^2],\\
    \chi_{t+1} &= \chi_t\mathbb{E}_{w_0,\xi}\qty[\eta'(\alpha w_0+\sqrt{\alpha\sigma^2_t}\xi;\chi_t)],
\end{split}
\end{equation}
where $\xi\sim\mathcal{N}(0,1)$ and $w_0$ obeys to a distribution of the true parameter.

\begin{figure}
    \centering
    \includegraphics[width=1.\linewidth]{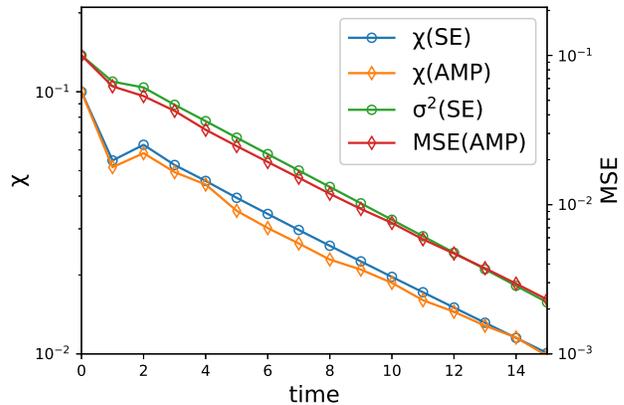}
\caption{Time evolution of $\chi$ on the left axis and of MSE on the right axis by SE (circles) and AMP (diamonds) for $\rho=0.1$, $\alpha=0.3$, and $\tau=0.4$ and the system size used in AMP is $N=10^4$.}
\label{fig:SE_AMP__rho0.1_alpha0.3_tau0.4}
\end{figure}

Fig.~\ref{fig:SE_AMP__rho0.1_alpha0.3_tau0.4} illustrates the time evolutions of AMP for a randomly generated instance and those of SE, which is the macroscopically averaged result. The time evolutions of the macroscopic quantities such as MSE and $\chi$  are in good agreement for SE and AMP. This allows us to discuss the nature of the convergent solutions of AMP by analyzing the fixed points of SE. In the statistical physics of random systems, the counterpart of SE is the saddle point equation for the RS free energy discussed below.

\subsection{Formulation in Statistical Physics}
From the viewpoint of Bayesian statistics, compressed sensing is defined as the MAP estimation of the parameters $\bm w$ and the local shrinkage parameter $\bm\lambda\in\mathbb{R}^N$ from the posterior distribution
\[
p(\bm w,\bm \lambda|\bm y,\bm X) = \frac{1}{Z(\beta|\bm y,\bm X)}\delta(\bm y - \bm X\bm w)e^{-\beta\mathcal{H}(\bm w,\bm\lambda)}, 
\]
where $\mathcal{H}(\bm w,\bm\lambda)$ satisfies
\[
\mathcal{H}(\bm w,\bm\lambda) \coloneqq \ln\prod^N_{i=1}\pi(\lambda_i)\mathcal{N}(w_i|0,\tau^2\lambda_i^2). 
\]
When the prior distribution for the local shrinkage parameter is the half-Cauchy prior given by $\pi(\lambda)=\frac{2}{\pi}\frac{1}{1+\lambda^2}$, the prior distribution for $\bm w$ given by marginalizing $\bm\lambda$ is called the horseshoe prior. Here, it is noted that $\{\lambda_i\}$, treated as hidden variables,  are determined by the MAP estimation as $\beta\rightarrow\infty$. Free entropy $\phi(\beta)$ is defined as an average of the logarithm of the normalization constant $Z(\beta|\bm{y},\bm{X})$, which is known as the partition function in physics, over the randomness of the true signal and the design matrix, expressed as 
\[
\phi(\beta) \coloneqq \frac{1}{N}\mathbb{E}_{\bm w_0,X}\qty[\ln Z(\beta|\bm y,\bm X)].
\]

In our study, the random variables $\bm{w}$ and $\bm{X}$ are generated from the Gaussian-Bernoulli distribution as 
\begin{equation}
    w_{0,i}\sim\rho\mathcal{N}(w_{0,i}|0,1) + (1-\rho)\delta(w_{0,i}),\nonumber
\end{equation}
and the Gaussian distribution as 
\begin{equation}
    X_{ij}\sim\mathcal{N}\left(X_{ij}|0,\frac{1}{N}\right).\nonumber
\end{equation}
Free energy is then defined as
\[
f(\beta) = -\frac{1}{\beta}\phi(\beta),
\]
and the limit of $\beta\rightarrow\infty$ is our interest. Using the replica method as in the spin glass theory\cite{mezard1987spin}, the average of the logarithm of the partition function is evaluated with the identity  
\begin{equation}\label{eqn:replica}
\qty[\ln Z(\beta|\bm y,\bm X)] = \lim_{n\rightarrow 0}\pdv{}{n}[Z^n(\beta,|\bm y,\bm X)],
\end{equation}
where the right hand size is explicitly calculated for $n\in\mathbb{N}^+$, taking the limit of $n\rightarrow0$ as the analytic continuation.

\section{Macroscopic dynamics of approximate message passing}
\label{sec:3}
\subsection{Saddle point equations for the replica symmetric free energy}
Assuming the replica symmetry condition, we obtain the RS free energy as 
\begin{equation}\label{eqn:RS_free_energy}
\begin{split}
-\frac{1}{\beta}\phi(\beta) = -\max_{q,\chi,m}\underset{\hat Q,\hat\chi,\hat m}{\mbox{extr}}&\hat\psi(\hat Q,\hat\chi,\hat m) + \alpha\psi(q,\chi,m)\\
~~~~&+ \frac{1}{2}(\hat Q q- \hat\chi\chi) - \hat m m,
\end{split}
\end{equation}
where 
\begin{equation}\label{eqn:hatpsi}
    \hat\psi = \mathbb{E}_{w_0,\xi}\qty[\max_{\lambda}\frac{(\hat m w_0 + \sqrt{\hat\chi}\xi)^2}{2(\hat Q + \tau^{-2}\lambda^{-2})} + \ln\pi(\lambda)],
\end{equation}
and 
\begin{equation}
    \psi = -\frac{\rho-2m+q}{2\chi}.
\end{equation}
The extremum condition for the RS free energy yields the saddle-point equations: 
\begin{equation}\label{eqn:saddle_point}
 \begin{split}
     q &= \mathbb{E}_{w_0,\xi}\qty[\left\langle\qty(\frac{w_0\hat m+\sqrt{\hat\chi}\xi}{\hat Q+\tau^{-2}\lambda^{-2}})^2\right\rangle],\\
    m &= \mathbb{E}_{w_0,\xi}\qty[\left\langle\frac{w_0^{2}\hat m+w_0\sqrt{\hat\chi}\xi}{\hat Q+\tau^{-2}\lambda^{-2}}\right\rangle],\\
    \chi &= \mathbb{E}_{w_0,\xi}\qty[\left\langle\frac{\xi^2+\frac{w_0\hat m}{\sqrt{\hat\chi}}\xi}{\hat Q+\tau^{-2}\lambda^{-2}}\right\rangle],\\ 
    \hat Q &= \alpha\frac{1}{\chi},\\ 
    \hat m &= \alpha\frac{1}{\chi},\\ 
    \hat\chi &= \alpha\frac{\rho-2m+q}{\chi^2}, 
\end{split}
\end{equation}
where $\langle\cdots\rangle$ denotes the thermal average defined as 
\begin{equation}
    \langle f(\lambda)\rangle = f(\lambda^*),\nonumber
\end{equation}
with 
\begin{equation}
    \lambda^* = \argmax_{\lambda}\frac{(\hat m w_0 + \sqrt{\hat\chi}\xi)^2}{2(\hat Q + \tau^{-2}\lambda^{-2})} + \ln\pi(\lambda).\nonumber
\end{equation}
Details of the derivation of the RS free energy are given in Appendix.~\ref{sec:appA}. 

Under the RS assumption, the MSE averaged with respect to the random variables is then expressed as 
\begin{equation}
    \mathrm{MSE} = \rho -2m +q. 
\end{equation}
Furthermore, the saddle point equations for the RS free energy are shown under a plausible assumption to be equivalent to the SE mentioned above. See Appendix~\ref{sec:app2} for details. 

\subsection{Linear Stability}
The stability of the true signal $w=w_0$ with respect to perturbations can be discussed using the linear stability of the solution of the saddle point equations characterized by $q=m=\rho$ and $\chi\rightarrow0$. By examining this limit, we obtain the following linear stability conditions:  
\begin{equation}\label{eqn:linearstability}
    \alpha > \rho+2(1-\rho)\Phi\qty(\sqrt{\frac{1}{\Delta}}), 
\end{equation}
where $\Phi$ is the cumulative distribution function of the normal Gaussian distribution and $\Delta$ is the solution of the following equation: 
\begin{equation}\label{eqn:Delta}
\begin{split}
    \alpha\Delta &= \rho\qty(\Delta+\qty[\frac{u(\lambda^*(w_0))}{u(0)}]_{w_0\sim\mathcal{N}(0,1)})\\
    &~~~~+2(1-\rho)\qty((1+\Delta)\Phi\qty(\sqrt{\frac{1}{\Delta}})-\sqrt{\frac{\Delta}{2\pi}}\exp(-\frac{1}{2\Delta}))\nonumber
\end{split}
\end{equation}
for
\begin{equation}
    u(\lambda) \coloneqq -\frac{(\ln\pi(\lambda))'}{\lambda}\nonumber, 
\end{equation}
and $\lambda^*$ satisfies
\begin{equation}
u(\lambda^*) \coloneqq \frac{w_0^2}{\tau^2(\lambda^*)^4}. \nonumber
\end{equation}
The derivation of this condition is provided in Appendix.~\ref{sec:app3}. The smallest $\alpha$ for which the true signal solution is stable under a given $\rho$ is of practical importance. We define $\alpha_{ls}$ as the smallest $\alpha$ that satisfies Eq.~(\ref{eqn:linearstability}) and will discuss it later. 

\subsection{Convergent Solution}
\label{sec:convergent_solution}
For successful recovery, the true solution must be stable, but this is not a sufficient condition. By examining the iterative dynamics and convergent solutions, one can identify regions where the recovery by the AMP algorithm can succeed. In particular, since the macroscopic time evolution of AMP is equivalent to an iterative substitution of the saddle point equations, the convergence condition to the iterative solution satisfying $\mathrm{MSE}=0$ corresponds to the successful recovery region of the true signal.

The iterative solution is obtained by repeating in the forward iterations of Eq.~(\ref{eqn:saddle_point}), which are saddle point equations of the order parameters, given by 
\begin{equation}
 \begin{split}
     q_t &= \mathbb{E}_{w_0,\xi}\qty[\left\langle\qty(\frac{\hat m_tw_0+\sqrt{\hat\chi_t}\xi}{\hat Q_t+\tau^{-2}\lambda^{-2}})^2\right\rangle],\\
    m_t &= \mathbb{E}_{w_0,\xi}\qty[\left\langle\frac{\hat m_t w_0^{2}+w_0\sqrt{\hat\chi_t}\xi}{\hat Q_t+\tau^{-2}\lambda^{-2}}\right\rangle],\\
    \chi_t &= \mathbb{E}_{w_0,\xi}\qty[\left\langle\frac{\xi^2+\frac{\hat m_t w_0}{\sqrt{\hat\chi_t}}\xi}{\hat Q_t+\tau^{-2}\lambda^{-2}}\right\rangle],\\ 
    \hat Q_{t+1} &= \alpha\frac{1}{\chi_t},\\ 
    \hat m_{t+1} &= \alpha\frac{1}{\chi_t},\\ 
    \hat\chi_{t+1} &= \alpha\frac{\rho-2m_t+q_t}{\chi_t^2}.
\end{split}
\label{eqn:iteration_SP}
\end{equation}
It should be noted that this time evolution is equivalent to SE under a suitable condition, as shown in Appendix \ref{sec:app2}, which allows us to evaluate the performance of AMP. 

\begin{figure}
    \centering
    \includegraphics[width=1.0\linewidth]{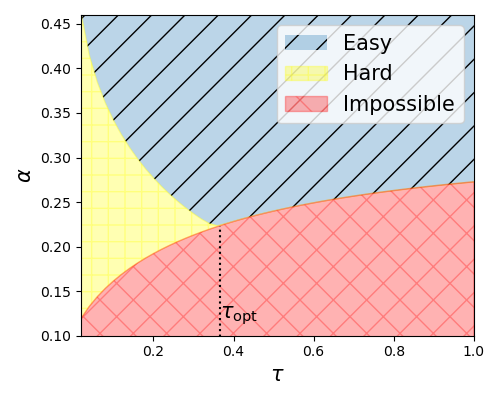}
    \caption{
    Phase diagram in $\tau-\alpha$ plane at $\rho=0.1$. In the Easy phase, the state evolution converges to a state of the true signal. In the Hard phase, the algorithm usually does not converge to this state because of strong initial condition dependence. In the Impossible phase below the linear stability line, the true signal is unstable. The value of $\tau$ that provides the minimum $\alpha$ value in the Easy phase is denoted b $\tau_\mathrm{opt}$.  
    }
    \label{fig:phase}
\end{figure}

The following two types of initial conditions are considered in applying AMP: 
\begin{itemize}
    \item non-informative: $q_0 = m_0 = 0$ and $\chi_0=10^{-1}$, 
    \item informative: $q_0 = m_0 = \rho - 10^{-6}$ and $\chi_0 = 10^{-3}$. 
\end{itemize}
The non-informative initial condition is $\mathrm{MSE}=\rho$, which may correspond to the initial state of AMP being the zero vector, while the informative condition is $\mathrm{MSE}\simeq 0$, which corresponds to the true vector. 
A phase diagram of compressed sensing using the AMP algorithm can be obtained by numerically solving the saddle point equations. As an example, 
Fig.~\ref{fig:phase} shows the phase diagram on the $\tau-\alpha$ plane at $\rho=0.1$. It can be seen that there are three distinct phases in common: Easy phase, Impossible phase, and Hard phase. In the Easy phase, the iterative solution converges to the true solution with $\mathrm{MSE}=0$, indicating that the AMP algorithms successfully recover the true signal. In the Impossible phase, the true signal is unstable, and then any algorithm which searches for a solution with the free-energy minimum fails to recover the true signal. In the Hard phase, the true signal is stable as in the Easy phase, but the AMP algorithms with an initial state of no information about the true signal converge to a solution with $\mathrm{MSE}\neq0$. For convergence to the true signal, an initial state close to the true signal is required. This situation is similar to that of the Bayes optimal case\cite{krzakala2012probabilistic}. The minimal value of $\alpha$ in the Easy phase is denoted by $\alpha_c(\tau)$, which is the limit where the true signal can be recovered from the non-informative initial condition for a fixed $\tau$.  

\begin{figure}
    \centering
    \includegraphics[width=1.0\linewidth]{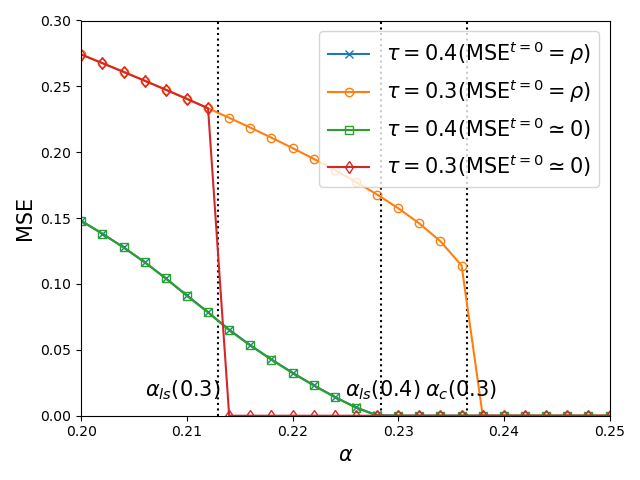}
    \caption{
    Converged mean-squared error (MSE) of the saddle-point equations as a function of $\alpha$ for $\rho=0.1$ and at $\tau=0.3$ and $0.4$. The results for non-informative and informative initial conditions at $\tau=0.4$ are represented by crosses and squares, while those for non-informative and informative conditions at $\tau=0.3$ are by diamonds and circles, respectively. 
    }
    \label{fig:FOT}
\end{figure}

Fig.~\ref{fig:FOT} shows $\alpha$ dependence of the MSE of the convergent solution of the saddle point equations starting from the two initial conditions, the informative and non-informative conditions.  At $\tau=0.4$, as $\alpha$ increases, the phase transition from the Impossible phase to the Easy phase occurs, which is a second-order transition. There is no dependence of converged MSE on the initial state at any $\alpha$. The second-order phase transition is supported by the finite-size simulations and their finite-size scaling analysis presented in Appendix.~\ref{app:D}. On the other hand, at $\tau=0.3$, as $\alpha$ increases, the phase transition from the Impossible phase to the Hard phase and then from the Hard phase to the Easy phase occurs, where the MSE behaves like a first-order transition, showing strong initial condition dependence. In the Hard phase, the converged MSE depends on the initial state and stays at $\mathrm{MSE}=0$ under the informative condition as long as the linear stability is satisfied, while it is trapped at a local minimum solution with $\mathrm{MSE}\neq0$ under the non-informative condition. The details of this picture are discussed  in Sec.\ref{sec:4}.


\begin{figure}
    \centering
    \includegraphics[width=1.0\linewidth]{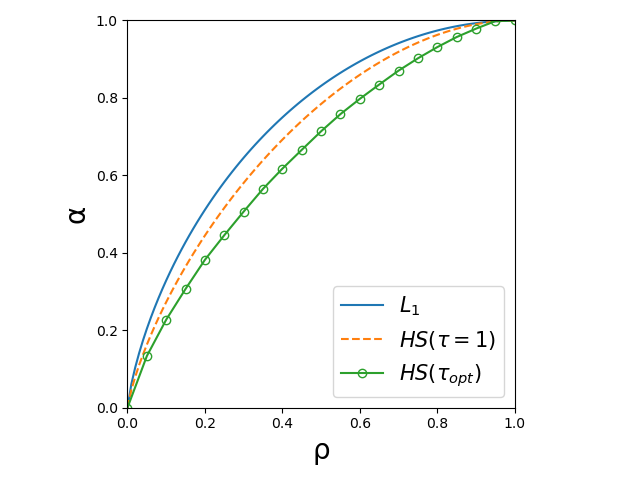}
    \caption{Phase diagram in $\rho-\alpha$ plane. Each line represents the boundary between successful and unsuccessful signal recovery, where $L_1$(filled) is for the $l_1$-norm regularization method, the two lines in HS are for the horseshoe prior method with different values of the global shrinkage parameter,  $\tau=1$(dashed) and $\tau_{\mathrm{opt}}$ (circle) with the optimal choice of $\tau$, respectively. ) is for no global shrinkage case. $\tau\rightarrow0$(green) is for linear stability and reaches $\alpha=\rho$. $\tau_{\mathrm{opt}}$(red) is for optimal choice of $\tau$. 
    The AMP algorithm can recover the true signal above the $\tau_{\mathrm{opt}}$ line. 
    }
    \label{fig:lasso_vs_HS_opt}
\end{figure}

\begin{figure}
    \centering
    \includegraphics[width=1.0\linewidth]{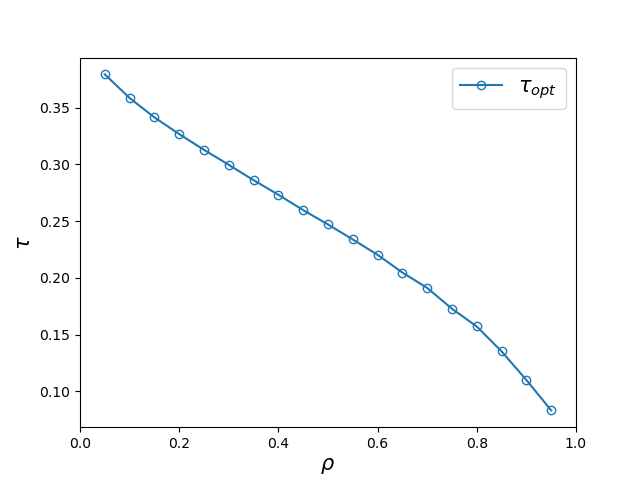}
    \caption{$\rho$ dependence of $\tau_{\mathrm{opt}}$ that gives the achievable bound $\alpha(\tau_{\mathrm{opt}})$ for the AMP algorithm. 
    }
    \label{fig:tau_opt}
\end{figure}

Fig.~\ref{fig:lasso_vs_HS_opt} is the phase diagram of successful and unsuccessful signal recovery in the $\rho-\alpha$ plane. The phase boundaries in the phase diagram differ depending on the estimation method used, such as the $l_1$-norm regularization or the horseshoe prior. The phase boundary by the horseshoe prior with the global shrinkage parameter $\tau=1$ was obtained in our previous study\cite{nagano2023phase}. 
The achievable boundary $\alpha_c(\tau_{\mathrm{opt}})$ for the AMP algorithm when $\tau$ is chosen optimally is obtained by analyzing the iterative solutions of the saddle point equations under the non-informative condition, while the optimal value of $\tau$ obtained numerically, as shown in Fig.~\ref{fig:tau_opt}, significantly depends on $\rho$.  
The boundary is well below that of the $l_1$-norm regularization for any $\rho$, which means that the signal can be recovered with less observed data using the horseshoe prior.  


\section{Free Energy landscape}\label{sec:4}
Analysis of the saddle point equations of RS free energy indicates the existence of the Hard phase in the phase diagram of the AMP algorithm for the compressed sensing with the horseshoe prior. In Sec.~\ref{sec:3}, it was found that the AMP algorithm with the non-informative initial condition fails to recover the true signal, although the solution of $\mathrm{MSE}=0$ is stable in the Hard phase. There, the definition of the phases depends on the initial state of the iterative algorithm and is thus based on a dynamical perspective. In this section, we provide the phase diagram from a static viewpoint characterized by a free-energy landscape, using a definition of the phases  by a free-energy landscape that does not have the arbitrariness such as the initial states of the dynamics. Because the AMP algorithm converges to the extremum of RS free energy, the landscape of the RS free energy is closely related to the recoverability limit of the AMP algorithm.

\subsection{RS Free Energy as a function of $\chi$ and $\hat\chi$}
According to the saddle-point equations of Eq.~(\ref{eqn:saddle_point}), which determine the RS free energy, we see that $\hat Q$ and $\hat m$ are determined only by $\chi$, and then $q$ and $m$ by $\chi$ and $\hat\chi$. Therefore, the RS free energy can be given formally only by $\chi$ and $\hat\chi$, taking extreme values of the other variables. 
As a result, the free-energy landscape as a function of $\chi$ and $\hat\chi$ is expressed as  
\begin{equation}\label{eqn:RS_free_energy_chi_hatchi}
\begin{split}
\frac{1}{\beta}\phi(\chi,\hat\chi) = \underset{q,m,\hat Q,\hat m}{\mbox{extr}}~&\hat\psi(\hat Q,\hat\chi,\hat m) + \alpha\psi(q,\chi,m)\\
~~~~&+ \frac{1}{2}(\hat Q q- \hat\chi\chi) - \hat m m. \nonumber
\end{split}
\end{equation}
 Because the RS free energy has a minimum with respect to $\chi$ and extremum with respect to $\hat\chi$, the number and location of saddle points in the free energy landscape are important to understand the behavior of the algorithm. 

\begin{figure}
    \centering
    \includegraphics[width=1.0\linewidth]{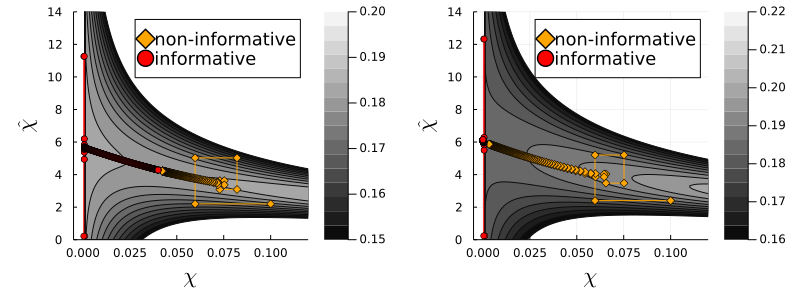}
    \caption{
    Free-energy landscape at $\rho=0.1$ and $\tau=0.4$ with $\alpha=0.22$ in the Impossible phase on the left and $\alpha=0.24$ in the Easy phase on the right. The lines with circles and diamonds represent respectively the dynamic trajectories of the iterative equation of Eq.~(\ref{eqn:iteration_SP}) for 100 steps starting from $(\chi,\hat{\chi})=(0.1,\alpha\rho/\chi^2)$ with the non-informative initial condition and for 200 steps from $(0.001,0.1)$ with the informative condition.  
    }
    \label{fig:contourf0.4}
\end{figure}

\begin{figure}
    \centering
    \includegraphics[width=1.0\linewidth]{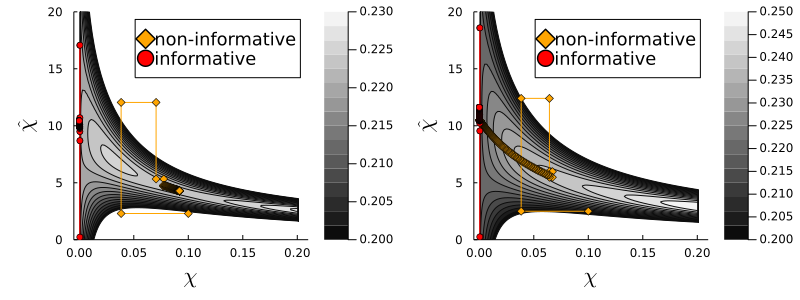}
    \caption{
    Free-energy landscape at $\rho=0.1$ and $\tau=0.3$ with $\alpha=0.23$ in the Hard phase on the left and $\alpha=0.25$ in the Easy phase on the right. The lines with circles and diamonds represent respectively the dynamic trajectories of the iterative equation of Eq.~(\ref{eqn:iteration_SP}) for 100 steps starting from $(\chi,\hat{\chi})=(0.1,\alpha\rho/\chi^2)$ with the non-informative initial condition and for 200 steps from $(0.001,0.1)$ with the informative condition. 
    }
    \label{fig:contourf0.3}
\end{figure}

Fig.~\ref{fig:contourf0.4} shows the free-energy landscape at $\rho=0.1$ and $\tau=0.4$ as an example of the phase transition from the Impossible phase to the Easy phase with $\alpha$ changing. In the figure, the lines represent the iterative dynamics of the iterative equations of Eq.~(\ref{eqn:iteration_SP}) for the two initial conditions, the informative and non-informative conditions, respectively. In the Impossible phase, both initial conditions converge to the saddle point where $\chi\neq 0$, while in the Easy phase, both converge to a saddle point with $\chi\rightarrow 0$ and $\hat{\chi}<\infty$, corresponding to the successful recovery. These are characteristic of the two phases.    
As another example of the phase transition from the Hard to Easy phase, Fig.~\ref{fig:contourf0.3} shows the free-energy landscape at $\rho=0.1$ and $\tau=0.3$. The Hard phase has a local maximum in the free-energy landscape at $(\chi,\hat{\chi})\simeq (0.04,7)$, which prevents the convergence of iterations from the non-informative condition to the successful recovery solution, while iterations from the informative condition converge stably to the successful solution. In the Easy phase, the local maximum still exists, but it is far from the two initial conditions and does not obstruct the convergence to the successful solution. This case implies that the MSE and $\chi$ converge discontinuously to zero with increasing $\alpha$.  

\begin{figure}
    \centering
    \includegraphics[width=1.0\linewidth]{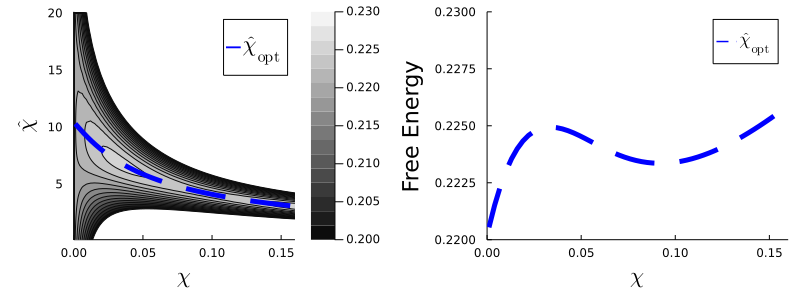}
    \caption{$\hat\chi_{\mathrm{opt}}$ on the free-energy landscape (left)  and the free energy along $\hat\chi_{\mathrm{opt}}$ as a function of $\chi$ (right)  at $\rho=0.1$, $\alpha=0.23$, and $\tau=0.3$, which is in the Hard phase.  }
    \label{fig:free_energy_as_function_of_chi}
\end{figure}

The above observations strongly suggest that the convergence solution is determined by the free energy on the curve $\hat{\chi}_{\mathrm{opt}}(\chi)$  in the free-energy landscape, defined by 
\begin{equation}
    f(\chi) = f(\chi,\hat\chi_{\mathrm{opt}}(\chi)),
\end{equation}
with
\begin{equation}
    \hat\chi_{\mathrm{opt}}(\chi) = \argmax_{\hat\chi}f(\chi,\hat\chi).
\end{equation}
Fig.~\ref{fig:free_energy_as_function_of_chi} shows the free-energy landscape on $\hat\chi_{\mathrm{opt}}$ at $\rho=0.1$, $\alpha=0.23$, and $\tau=0.3$, which is in the Hard phase. The non-monotonic landscape implies that the local search minimization of the free energy with respect to $\chi$ fails to converge to $\chi \to 0$, meaning that the AMP algorithm cannot recover the true signal although $\chi=0$ is a stable solution. 

\begin{figure}
    \includegraphics[width=1.0\linewidth]{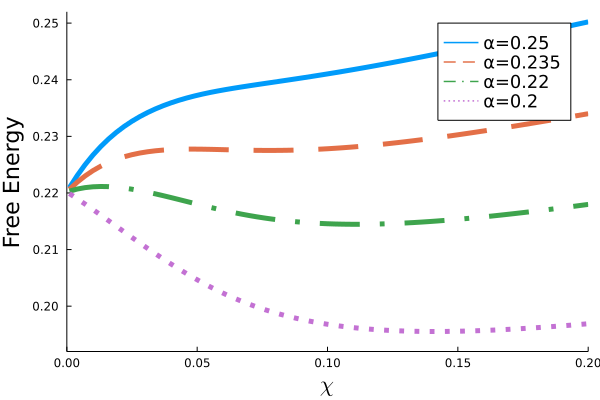}
    \caption{
    Free energy $f(\chi)$ in the three phases shown in Fig.~\ref{fig:phase} at four different values of $\alpha$, 0.25, 0.235, 0.22, and 0.2, corresponding to the Easy phase, the Hard-possible phase, the Hard-impossible phase, and the Impossible phase. See the main text for the definition of  
 these phases.   
    }
    \label{fig:landscape_and_phase}
\end{figure}

Fig.~\ref{fig:landscape_and_phase} shows the $\chi$ dependence of the free energy $f(\chi)$ in the three phases defined from the convergence conditions of the AMP algorithm in Sec.~\ref{sec:convergent_solution}. One can see a characteristic behavior in the $\chi$ dependence of the free energy. From this, we redefine the phases by this static characterization as follows: 
\begin{itemize}
    \item Easy phase: $f(\chi)$ is a monotonically increasing function of $\chi$
    \item Hard phase: $f(\chi)$ is not a monotonically function of $\chi$ and $f'(\chi)|_{\chi\to0}\geq0$
    \begin{itemize}
        \item Hard-possible phase: $f(0)$ is a global minimum
        \item Hard-impossible phase: $f(0)$ is not a global minimum
    \end{itemize}
    \item Impossible phase: $f(\chi)$ is not a monotonically function of $\chi$ and $f'(\chi)|_{\chi\to0}<0$
\end{itemize}
The Hard phase has been divided into two phases by the property of the global minimum of the free energy $f(\chi)$. If a global search could be performed, it would be possible to recover the true signal even in the Hard possible phase.  

\begin{figure}
    \centering
    \includegraphics[width=1.0\linewidth]{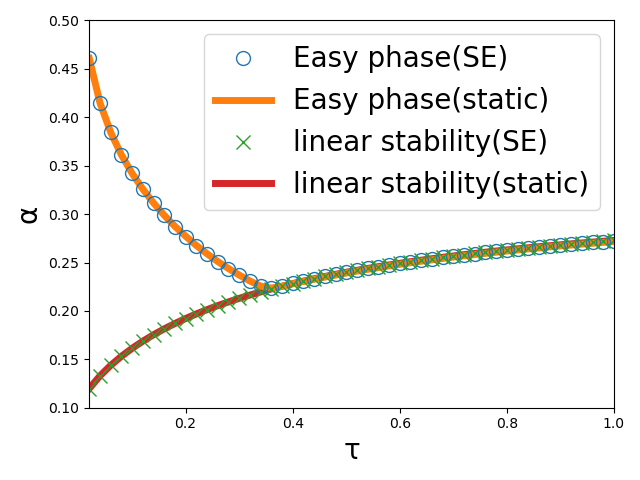}
    \caption{
    Phase diagram in $\tau-\alpha$ plane at $\rho=0.1$, the same as Fig.~\ref{fig:phase}, with static phase boundaries marked by lines and dynamics phase boundaries of the AMP algorithm by circles and crosses.  
    }
    \label{fig:landscape_vs_SPE}
\end{figure}

Since the monotonicity of the free energy landscape along $\hat\chi_{\mathrm{opt}}(\chi)$ is defined by static characterization, the definition of these phases has no arbitrary such as the initial conditions for the time evolution of the algorithms. However, the correspondence between the monotonicity of the free energy and the phases is highly non-trivial, since the iterative updates of the algorithm are not necessarily local searches. 
Therefore, we compare the static phase boundaries based on the monotonicity of the free-energy landscape with the dynamic phase boundaries characterized by the convergence value of the AMP algorithm. As shown in Fig.~\ref{fig:landscape_vs_SPE}, it demonstrates that the two boundaries coincide with each other over a wide range,  suggesting that in the Hard phase, the AMP algorithm does indeed converge to a local minimum of the free energy. 
The correspondence of phase boundaries always holds for $0<\rho<1$. 

\subsection{Path of damped dynamics}
In the previous subsection, the static characterization of the phases is given by the monotonicity of the free-energy landscape along the optimal curve $\hat\chi_{\mathrm{opt}}$, where the free-energy landscape is optimized first with respect to $\hat\chi$, and then with respect to $\chi$. On the other hand, if $\chi$ is first and $\hat\chi$ second, another optimal curve $\chi_{\mathrm{opt}}(\hat{\chi})$ is obtained as
\begin{equation}
    \chi_{\mathrm{opt}}(\hat\chi) = \argmin_{\chi}f(\chi,\hat\chi).
\end{equation}
We discuss the meaning of this curve in relation to the dynamics of SE. The iterative update of $\chi$ with a damping term in SE is given by  
\begin{equation}
    \chi_t = r_{\chi}\chi_{t-1}+(1-r_\chi)\chi(\hat Q_t,\hat\chi_t,\hat m_t), 
\end{equation}
where $r_\chi$ is a parameter representing the strength of the damping. When updating $\chi$ with $r_\chi>0$, the extreme condition with respect to $\hat{\chi}$ is not satisfied except at fixed points, while the extreme condition of $\chi$ is satisfied when updating $\hat{\chi}$. 
In the limit of the damping parameter $r_\chi\to1$, $\chi_t$ moves very slowly, and $\hat{\chi}$ also moves slowly. Note that only the extreme condition for $\chi$ is satisfied by the update of $\hat{\chi}$, indicating that SE with the heavy damping parameter moves along $\chi_{\mathrm{opt}}$. 

\begin{figure}
    \centering
    \includegraphics[width=0.9\linewidth]{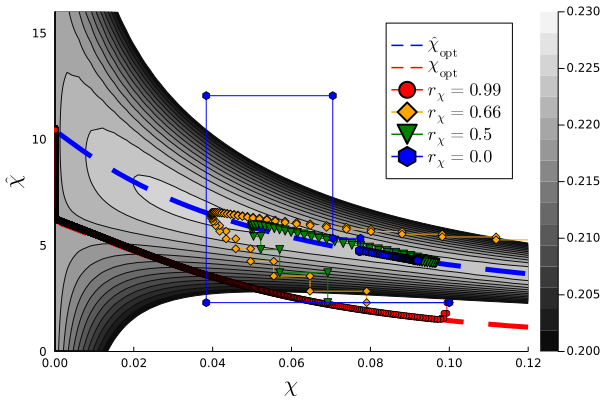}
    \caption{
    Free-energy landscape at $\alpha=0.23, \tau = 0.3$ and $\rho=0.1$ and the trajectory of the SE dynamics with several values of $r_\chi$. The two dashed lines are the optimal curves $\chi_{\mathrm opt}$ and $\hat{\chi}_{\mathrm opt}$, which optimized either of the two arguments of the free energy $f(\chi,\hat{\chi})$. 
    }
    \label{fig:various_damp_optimalline}
\end{figure}

Fig.~\ref{fig:various_damp_optimalline} shows trajectories of the SE dynamics with some damping parameters and the two optimal curves, $\chi_{\mathrm{opt}}$ and $\hat{\chi}_{\mathrm{opt}}$ on the free-energy landscape with $\alpha=0.23$ and $\rho=0.1$, which is in the Hard phase. It can be seen that the iterative dynamics converges to the local minimum of $\hat{\chi}_{\mathrm opt}$ for $r_\chi=0$, and behaves in a complicated manner when $r_\chi$ is finite and not very large. However,  when $r_\chi=0.99$, which is heavy damping, the trajectory of the iterative dynamics moves very slowly along $\chi_{\mathrm{opt}}$ and eventually reaches $\chi\to 0$. This means that even in the Hard phase where there exists a local minimum on $\hat{\chi}_{\mathrm opt}$, the dynamics with damping parameters moving along the path $\chi_{\mathrm{opt}}$ may successfully estimate the true signal. In this case, a necessary condition for success is that the optimal curve $\chi_{\mathrm{opt}}$ reaches zero. 

\begin{figure}
    \centering
    \includegraphics[width=0.9\linewidth]{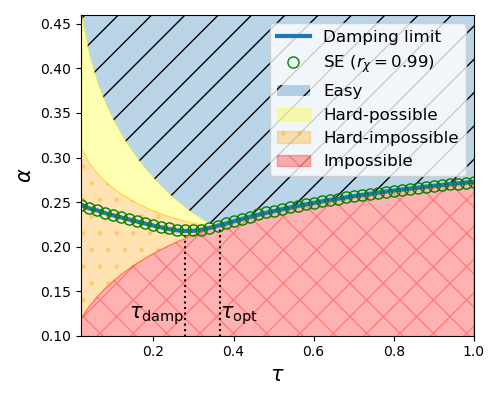}
    \caption{
    Phase diagram in $\tau-\alpha$ plane at $\rho=0.1$. The thick solid line is the damping phase boundary characterized by $\chi_{\mathrm{opt}}$, and circles represent the convergence limit to the true signal by SE with $r_\chi=0.99$. The global parameter that gives the minimum successfully recovered $\alpha$ is denoted $\tau_{\mathrm{damp}}$ and $\tau_{\mathrm{opt}}$ with and without damping, respectively.}
    \label{fig:phase_via_landscape_vs_algorithm}
\end{figure}

Here, the additional phase is defined as the region where $\chi_{\mathrm{opt}}$ reaches zero and is called the Damping phase, as distinguished from the Hard-possible and Hard-impossible phases. Fig.~\ref{fig:phase_via_landscape_vs_algorithm} presents the phase boundary of the Easy phase and of the Damping phase, showing that the reconstruction limit can be extended by appropriately choosing $\tau$ and by using heavy damping. We also plot the recoverable boundaries of the SE dynamics with heavy damping parameters, which is in good agreement with the Damping phase boundary. It should be noted that the Damping phase is not characterized by the saddle point property of free energy and is therefore different in nature from the phase of an equilibrium system in the physics sense. As seen in Fig.~\ref{fig:phase_via_landscape_vs_algorithm}, it is interesting to notice that the reconstruction limit line with the heavy damping is significantly below the phase boundary between the Hard-possible and Hard-impossible phases, indicating that the damping has better estimation performance than the global search.  

\section{Summary and Discussion}\label{sec:5}
One of the central issues in the context of the method of the horseshoe prior distribution has been to uncover the role of the global shrinkage parameter in estimation accuracy. In this paper, we defined compressed sensing using a horseshoe prior and revealed the effect of the global-shrinkage parameter $\tau$ on the phase diagram through statistical physics analysis. According to the phase diagram, the local stability of the true signal improves with decreasing $\tau$ in terms of the linear stability boundary. However, for $\tau < \tau_{\rm{opt}}$, there exists the Hard phase, in which the algorithm converges to a fixed point other than the true signal even though the true signal solution is linearly stable. This makes the dependence of the algorithmic phase transition point $\alpha_c(\rho,\tau)$ on $\tau$ nonmonotonic. Furthermore, such dynamics of the algorithm is found to be explained by whether or not there is a local fixed point in the free-energy landscape. Then, the phase diagram obtained from the dynamics is successfully characterized statically by the free energy landscape. This structure is similar to the picture of the Bayes optimal case.

There remains much work to be conducted on the dependence of optimization problems on the solving algorithm. Our analysis has demonstrated that there exists a region in the Hard phase where the true signal can be efficiently recovered by adding  a damping term to the AMP algorithm, which slightly broadens the algorithmic limit for the recovery of the sparse signal. In particular, we emphasize the importance of the path followed by the dynamics, in addition to the free-energy saddle point, in comprehending the performance of the algorithmic reconstruction from a macroscopic viewpoint, although this has not previously received much attention. To confirm the existence of the path to the true solution that avoids local solutions, it is necessary to examine a free-energy landscape expressed in at least two variables, suggesting that the conventional discussion of a free-energy landscape with only the mean squared error as a variable is insufficient. It is also confirmed numerically that the free-energy landscape can explain the behavior of the standard AMP algorithm without damping on the optimal curve $\hat\chi_{\rm{opt}}$, although this is not theoretically obvious due to the discontinuous updates. While it is useful to understand optimization problems statically using the free-energy landscape, it should be well noted that the dynamics of the algorithm is characterized not only by the number of saddle points in the free-energy landscape but also by a more global structure.

\begin{figure}
    \centering
    \includegraphics[width=0.9\linewidth]{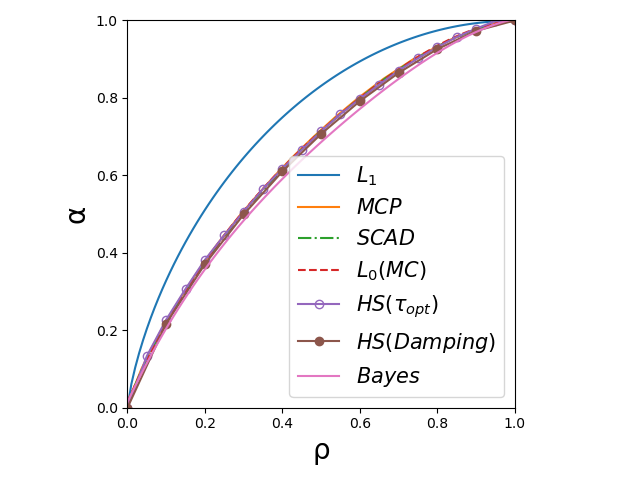}
    \caption{
    Reconstruction-limit line as a function of $\rho$  in several algorithms.
    The meaning of each abbreviation is as follows: $L_1$ is the $\ell_1$-norm regularization. MCP and SCAD are both non-convex sparse penalty-based methods, Minimax Concave Penalty and Smoothly Clipped Absolute Deviation, respectively. $L_0$(MC)\cite{obuchi2018statistical} is the $\ell_0$ norm regularization with Monte Carlo search. ``Bayes'' is the Bayes optimal reconstruction. The two HS are the optimal lines in the horseshoe prior method with $\tau_{\mathrm{opt}}$ and the damping. 
    }
    \label{fig:compare_reconstruction_limit}
\end{figure}

Eventually, reconstruction limits for several existing algorithms are shown in Fig.~\ref{fig:compare_reconstruction_limit}. Except for the $\ell_1$-norm regularization corresponding to the Laplace prior, and the Bayesian optimal case where the generating distribution and the prior coincide, the promising methods exhibit nearly equivalent recovery performance. As with other methods such as SCAD, the horseshoe prior method is efficient in the sense that its performance is similar to the Bayes optimal reconstruction even without assuming prior knowledge of the true signal. One of the main results of this study is to clarify the efficiency of the turning of the global parameter. This suggests that compressed sensing utilizing the horseshoe prior is one of the most promising methods, although the relative superiority of the methods is expected to depend on the specific generative distribution.

\begin{acknowledgments}
We would like to thank T.~Obuchi and A.~Sakata for the fruitful discussions and for providing numerical data for Fig.~\ref{fig:compare_reconstruction_limit}. 
This work was supported by MEXT as the Program for Promoting Research on the Supercomputer Fugaku (DPMSD, Project ID: JPMXP1020200307). One of the authors, YN, was supported by the SPRING-GX program at the University of Tokyo.
\end{acknowledgments}

\bibliography{reference} 
\bibliographystyle{unsrt} 
\appendix

\section{Derivation of Replica Symmetric Free Energy}
\label{sec:appA}
Here, we explain the derivation of the replica symmetric (RS) free energy in Eq.~(\ref{eqn:RS_free_energy}), which is  almost the same as in our previous paper\cite{nagano2023phase}.
According to Eq.~(\ref{eqn:replica}), the free energy is obtained by the replicated partition function $[Z^n]$ with the true signal $\bm{w}_0$, which is given by  
\begin{equation}
    [Z^n] = \qty[\int\qty(\prod^n_{a=1}\dd{w_a}\delta(\bm X(\bm w_0- \bm w_a))\dd{\lambda_a}\pi(\lambda_a))e^{-\beta\sum^n_{i=1}\mathcal{H}(\bm w_a\bm\lambda_a)}].
\end{equation}
Since $X_{i,j}$ is a Gaussian variable following the i.i.d Gaussian distribution $\mathcal{N}(0,N^{-1})$, $y^i_a\coloneqq\bm x_i^\top\bm w_a$ can also be treated as a Gaussian variable. Order parameters $m_a$ and $q_{ab}$ are defined as elements of the variance-covariance matrix for $\bm{y}_a$ as 
\begin{equation}\label{eqn:order_parameters}
\begin{split}
    \mathbb{E}_{\bm x}[y_0y_0] &= \frac{1}{N}\bm w_0^\top\bm w_0 = \rho,\\
    \mathbb{E}_{\bm x}[y_0y_a] &= \frac{1}{N}\bm w_0^\top\bm w_a \eqqcolon m_a,\\
    \mathbb{E}_{\bm x}[y_ay_b] &= \frac{1}{N}\bm w_a^\top\bm w_b \eqqcolon q_{ab}.\nonumber
\end{split}
\end{equation}
Inserting these order parameters as constraints with Lagrange multipliers for $[Z^n]$, we obtain
\begin{equation}\label{eqn:Zn}
\begin{split}
[Z^n] &= \underset{\{\tilde q_{ab},\tilde m_a\}}{\mathrm{extr}}\qty(\int_{\mathbb{R}^{n+1}}\dd{\bm y}\mathcal{N}(\bm y|\bm 0,\bm \Sigma)\prod^n_{a=1}\delta(y_0-y_a))^N\\
&~~\cdot\int\prod^n_{a=1}\dd{w_a}\exp(-\beta\sum^N_{i=1}\sum^n_{a=1}\mathcal{H}(w_{a,i}))\\
&~~\cdot\exp(\sum_{a\leq b}\tilde q_{ab}(Nq_{ab}-\sum^N_{i=1}w_{a,i}w_{b,i}))\\
&~~\cdot\qty[\exp(\sum_{a}\tilde m_{a}(Nm_{a}-\sum^N_{i=1}w_{0,i}w_{a,i}))]_{\bm w_0}, 
\end{split}
\end{equation}
where
\begin{equation}\label{eqn:Sigma}
    \bm\Sigma = \mqty(\rho && \bm m^\top \\ \bm m && \bm Q)\in \mathbb{R}^{n+1},(\bm Q)_{ab} = q_{ab}, (\bm m)_a = m_a.
\end{equation}
Under the replica symmetric assumption, the order parameters and corresponding conjugate parameters have no dependence on the replica index as 
\begin{equation}\label{eqn:RS}
\begin{split}
    q_{aa} &= q+\frac{\chi}{\beta},\\
    q_{ab} &= q,\\
    m_a &= m,
\end{split}
\end{equation}
and 
\begin{equation}\label{eqn:RSconjugate}
\begin{split}
\tilde q_{aa} &= \beta\hat Q -\beta^2\hat\chi,\\
\tilde q_{ab} &= -\beta^2\hat\chi,\\
\tilde m_a &= -\beta\hat m.\\
\end{split}
\end{equation}
By substituting Eq.~(\ref{eqn:RS}) and (\ref{eqn:RSconjugate}) into (\ref{eqn:Zn}) and (\ref{eqn:Sigma}),  using Hubbard-Stratonovich transformation
\begin{equation}
    e^{\beta^2\hat\chi(\sum_{a}w_a)^2} = \int D\xi~e^{\beta\sqrt{\hat\chi}\sum_a{w_a}\xi},
\end{equation}
and taking the limits $N\rightarrow\infty,~n\rightarrow0$ and $\beta\rightarrow\infty$, 
the RS free energy is obtained as Eq. (\ref{eqn:RS_free_energy})

\section{Equivalence between saddle point equations and state evolution}\label{sec:app2}
We show here that the iteration equations for the saddle point equations for the RS free energy and the state evolution (SE) are equivalent. 
The iteration equations for solving the saddle point equations for the RS free energy are given by 
\begin{equation}
 \begin{split}
    \hat Q_{t+1} &= \alpha\frac{1}{\chi_t}, \\ 
    \hat m_{t+1} &= \alpha\frac{1}{\chi_t},\\ 
    \hat\chi_{t+1} &= \alpha\frac{\rho-2m_t+q_t}{\chi_t^2},\\
    q_t &=\mathbb{E}_{w_0,\xi}\qty[\left\langle\qty(\frac{w_0\hat m_t+\sqrt{\hat\chi_t}\xi}{\hat Q_t+\tau^{-2}\lambda^{-2}})^2\right\rangle],\\
    m_t &= \mathbb{E}_{w_0,\xi}\qty[\left\langle\frac{w_0^{2}\hat m_t+w_0\sqrt{\hat\chi_t}\xi}{\hat Q_t+\tau^{-2}\lambda^{-2}}\right\rangle],\\
    \chi_t &= \mathbb{E}_{w_0,\xi}\qty[\left\langle\frac{\xi^2+\frac{w_0\hat m_t}{\sqrt{\hat\chi_t}}\xi}{\hat Q_t+\tau^{-2}\lambda^{-2}}\right\rangle].
    \nonumber
\end{split}
\end{equation}
From these equations, the iterative dynamics of the mean squared error $\sigma^2$ and the variance $\chi$ are derived in the following. First, substituting $q_t$ and $m_t$ into the mean squared error $\sigma^2_t\coloneqq\rho-2m_t+q_t$, we have the iteration equation for $\sigma_t$ as 
\begin{equation}\label{eqn:sigma}
\begin{split}
    \sigma^2_t &= \mathbb{E}_{w_0,\xi}\qty[\left\langle\frac{w_0\hat m_t+\sqrt{\hat\chi_t}\xi}{\hat Q_t+\tau^{-2}\lambda^{-2}} - w_0\right\rangle^2],\\
    &=\mathbb{E}_{w_0,\xi}\qty[\left\langle\frac{\alpha w_0+\sqrt{\alpha\sigma_{t-1}^2}\xi}{\alpha+\chi_{t-1}\tau^{-2}\lambda^{-2}} - w_0\right\rangle^2],\\
    &=\mathbb{E}_{w_0,\xi}\qty[\left(\eta(h_{t-1};\chi_{t-1},\alpha,\tau) - w_0\right)^2], 
\end{split}
\end{equation}
where $\eta$ is the threshold function and $h_t\coloneqq\alpha w_0 + \sqrt{\alpha\sigma^2_{t}}\xi$. Similarly, the iteration equation for $\chi_t$ is obtained as 
\begin{equation}\label{eqn:chi}
\begin{split}
    \chi_t &= \mathbb{E}_{w_0,\xi}\qty[\left\langle\frac{\xi^2+\frac{w_0\hat m_t}{\sqrt{\hat\chi_t}}\xi}{\hat Q_t+\tau^{-2}\lambda^{-2}}\right\rangle],\\ 
    &=\mathbb{E}_{w_0,\xi}\qty[\frac{\xi}{\sqrt{\hat\chi_t}}\left\langle\frac{\hat m_t w_0 + \sqrt{\hat\chi_t}\xi}{\hat Q_t+\tau^{-2}\lambda^{-2}}\right\rangle],\\
    &=\mathbb{E}_{w_0,\xi}\qty[\frac{1}{\sqrt{\hat\chi_t}}\pdv{\xi}\left\langle\frac{\hat m_t w_0 + \sqrt{\hat\chi_t}\xi}{\hat Q_t+\tau^{-2}\lambda^{-2}}\right\rangle],\\
    &=\mathbb{E}_{w_0,\xi}\qty[\sqrt{\frac{\chi_{t-1}^2}{\alpha\sigma_{t-1}^2}}\pdv{\xi}\left\langle\frac{\alpha w_0 + \sqrt{\alpha\sigma_{t-1}^2}\xi}{\alpha+\chi_{t-1}\tau^{-2}\lambda^{-2}}\right\rangle],\\
    &=\chi_{t-1}\mathbb{E}_{w_0,\xi}\qty[\pdv{h}\eta(h_{t-1};\chi_{t-1},\alpha,\tau)].
\end{split}
\end{equation}
Eq.~(\ref{eqn:sigma}) and (\ref{eqn:chi}) are the deterministic dynamics in SE. 
From the second to the third line in Eq.~(\ref{eqn:chi}), a partial integral on $\xi$ is calculated. When $\eta$ is a discontinuous function, the partial integral is invalid. Therefore, the saddle point equation is equivalent to SE when $\tau^2\frac{\alpha}{\chi}>0.5$, which is the continuity condition for $\eta$.
Note that some differences in the coefficients from the original paper\cite{donoho2010message_2} are due to the normalization of the design matrix. 

\section{Linear Stability of true signal}\label{sec:app3}
In the statistical-mechanics formulation, the true signal estimation, i.e., $\hat {\bm w} \rightarrow \bm w_0$, implies $\chi\rightarrow0$, where $q\rightarrow \rho$, $m\rightarrow \rho$ and $\hat Q\rightarrow\infty$ are satisfied.  In this limit, only $\hat\chi$ has a non-trivial value, satisfying the relation 
\begin{equation}\label{eqn:hatchi}
\begin{split}
    \hat\chi &= \alpha\frac{(\rho-2m+q)}{\chi^2}\\
    &=\alpha\qty[\left\langle\frac{\hat m w_0+\sqrt{\hat\chi}\xi}{\hat Q + \tau^{-2}\lambda^{-2}}-w_0\right\rangle^2],\\
    &=\frac{1}{\alpha}\left(\rho\qty(\hat\chi+\qty[\frac{ u(\lambda^*(w_0))}{\tau^{2}}]_{w_0\sim\mathcal{N}(0,1)} ) \right.,\\
    &~~~~\left.+ 2(1-\rho)\qty[\qty(\sqrt{\hat\chi}\xi-\sqrt{\tau^{-2}u_0})^2]_{\xi>\sqrt{\frac{u_0}{\tau^2\hat\chi}}}\right).
\end{split}
\end{equation}
Eq.~(\ref{eqn:Delta}) was obtained by Eq.~(\ref{eqn:hatchi}) with the scaling $\Delta\coloneqq\frac{\tau^2\hat\chi}{u_0}$ and the linear stability condition is derived by the stability of  Eq.~(\ref{eqn:Delta}) with respect to perturbation $\Delta^*\rightarrow\Delta^*+\delta\Delta$.

\section{Numerical experiment: Finite Size Scaling for the second-order transition from the Impossible to the Easy phase.}
\label{app:D}
\begin{figure}[h]
    \centering
    \includegraphics[width=1.0\linewidth]{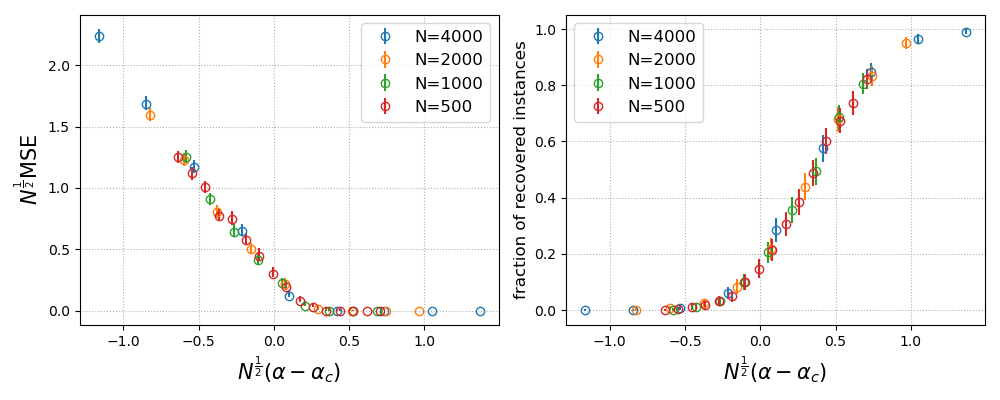}
    \caption{Finite size scaling plots of the mean squared error MSE (left) and the fractions of recovered instances(right) for $10^3$ instances for various system sizes at $\rho=0.1$ and $\tau=0.4$. Both scaling plots assume $\alpha_c=0.2284$.  }
    \label{fig:recover}
\end{figure}

In this appendix, we present the numerical results with finite size $N$ for the phase transition between the Impossible phase and the Easy phase. 
Fig.~\ref{fig:recover} shows the finite size scaling plots of MSE and recoverability, assuming $\alpha_c=0.2284$,  averaged over $10^3$ instances solved by AMP at $\rho=0.1$ and $\tau=0.4$,  a near-optimal $\tau$. When evaluating the recoverability, instances with MSE $<10^{-6}$ are regarded as successfully recovered. In our previous study\cite{nagano2023phase}, the phase transition point of the compressed sensing was evaluated as $\alpha_c\simeq 0.27$ at $\rho=0.1$ and $\tau=1.0$ for the purely local shrinkage case. The theoretical analysis in this study revealed that a more efficient recovery is possible for the global shrinkage $\tau<1$. In fact, from Fig.~\ref{fig:phase}, the second-order transition from the Impossible phase to the Easy phase occurs at $\alpha_c\simeq0.2284$, which is consistent with our numerical analysis.

\end{document}